\documentclass[aps, prd, showpacs, twocolumn, superscriptaddress, nofootinbib]{revtex4}
\usepackage{graphicx}
\usepackage{dcolumn}
\usepackage{amssymb,amsmath}
\usepackage{relsize}

\begin{document}

\newcommand{\Eq}[1]{\mbox{Eq. (\ref{eqn:#1})}}
\newcommand{\Fig}[1]{\mbox{Fig. \ref{fig:#1}}}
\newcommand{\Sec}[1]{\mbox{Sec. \ref{sec:#1}}}

\def\d{\mathrm{d}}

\def\mupn{^\mu_{\, \nu}}
\def\nn{\nonumber}

\def\K{\mathcal{K}}

\def\({\left(}
\def\){\right)}
\def\ie{{\it i.e. }}
\def\ms{M_6^4{}}
\def\mf{M_5^3{}}
\def\mq{M_4^2{}}
\def\pp{\partial_\mu\partial_\nu \pi}
\def\dbi{1+(\pa \pi)^2}
\def\tg{\tilde\gamma}

\newcommand{\detgt}{\sqrt{\tilde{g}}}
\newcommand{\detg}{\sqrt{g}}
\newcommand{\detgg}{\sqrt{\frac{\tilde{g}}{g}}}
\newcommand{\nabt}{\tilde{\nabla}}
\newcommand{\pa}{\partial}
\newcommand{\tmnt}{\tilde{T}^{\mu\nu}}
\newcommand{\tmn}{T^{\mu\nu}}
\newcommand{\mett}{\tilde{g}_{\mu\nu}}
\newcommand{\met}{g_{\mu\nu}}
\newcommand{\cS}{{\cal{S}}}
\newcommand{\cL}{{\cal{L}}}
\newcommand{\cE}{{\cal{E}}}
\newcommand{\HRule}{\rule{\linewidth}{0.4mm}}
\newcommand{\comment}[1]{}
\newcommand{\lagm}{{\cal{L}}_m}
\newcommand{\lage}{{\cal{L}}_E}

\def\O{{\cal O}}
\def\L{{\cal L}}
\def\mpl{M_{Pl}}

\newcommand{\ba}{\begin{eqnarray}}
\newcommand{\ea}{\end{eqnarray}}
\newcommand{\mn}{_{\mu\nu}}
\newcommand{\umn}{^{\mu\nu}}
\newcommand{\cA}{{\cal A}}
\newcommand{\dof}{{\it dof} }
\newcommand{\eft}{{\it EFT} }

\newcommand{\eref}{\eqref}

\newcommand{\PHI}{\phi}
\newcommand{\PhiN}{\Phi^{\mathrm{N}}}
\newcommand{\vect}[1]{\mathbf{#1}}
\newcommand{\Del}{\nabla}
\newcommand{\unit}[1]{\;\mathrm{#1}}
\newcommand{\x}{\vect{x}}
\newcommand{\ScS}{\scriptstyle}
\newcommand{\ScScS}{\scriptscriptstyle}
\newcommand{\xplus}[1]{\vect{x}\!\ScScS{+}\!\ScS\vect{#1}}
\newcommand{\xminus}[1]{\vect{x}\!\ScScS{-}\!\ScS\vect{#1}}
\newcommand{\diff}{\mathrm{d}}

\newcommand{\be}{\begin{equation}}
\newcommand{\ee}{\end{equation}}
\newcommand{\bea}{\begin{eqnarray}}
\newcommand{\eea}{\end{eqnarray}}
\newcommand{\vu}{{\mathbf u}}
\newcommand{\ve}{{\mathbf e}}

        \newcommand{\vU}{{\mathbf U}}
        \newcommand{\vN}{{\mathbf N}}
        \newcommand{\vB}{{\mathbf B}}
        \newcommand{\vF}{{\mathbf F}}
        \newcommand{\vD}{{\mathbf D}}
        \newcommand{\vg}{{\mathbf g}}
        \newcommand{\va}{{\mathbf a}}


\title{Screening Solutions in Modified Gravity Theories}

\newcommand{\addressImperial}{Theoretical Physics, Blackett Laboratory, Imperial College, London, SW7 2BZ, United Kingdom}
\newcommand{\addressOxford}{??, United Kingdom}

\author{Ali Mozaffari}
\email{ali.mozaffari05@imperial.ac.uk}
\affiliation{\addressImperial}

\date{\today}

\begin{abstract}
In this work, we illustrate through a simple example the possibility of testing the chameleon screening mechanism in the Solar System using the forthcoming LISA Pathfinder mission around gravitational saddle points.  We find distinctive tidal stress signatures for such models and consider the potential for constraints.
\end{abstract}


\maketitle

\subsection{Introduction}
In this short letter we aim to show that {\it chameleonesque} screening mechanisms
are naturally present in preferred acceleration modified gravity theories.
 As such they are of significant interest if we are looking to hide additional effects on Solar System scales and also tests of these modified theories should reach the level of potentially certifying or vilifying models
and classes of theories in the very near future~\cite{alireview}.  The use
of the forthcoming LISA Pathfinder mission~\cite{LPF2} provides an opportunity
to directly probe an unexplored low acceleration regime as well, as well
as test some less conventional ideas in gravitational physics~\cite{ali,Clifton11}.  This presents
the opportunity to probe the {\it onset} of such mechanisms and placing constraints
on them in a new and interesting way would be a key result.  At the saddle point (SP) of the gravitational field, we find the regime where
both $\rho \rightarrow 0$ and $|\Del \Phi_N| \rightarrow 0$ which we see is reminiscent of the ``cosmological regime''. 

We can consider as a simple example the Chameleon~\cite{khouryetal}, \bea S &=& \int \left(\frac{M^2_{pl}}{2}
R + X - V(\phi)\right)\sqrt{-g}\, d^4 x \nonumber \\&+& \int \mathcal{L}_m(\Psi_i) \sqrt{-\tilde{g}} \,d^4 x \\ \tilde{g}\mn &=& A^2(\phi)\, g\mn \\X &=& -\frac{1}{2}\Del_\mu\phi\Del^\mu\phi\\{\bf EoM:}\nonumber\\  \square \phi &=& V_{,\phi} - A^3(\phi) A_{,\phi} \tilde{T} = V_{eff,\phi}\eea where we find that $\tilde{T} = -\tilde{\rho}$ and $\tilde{\rho} = A^{-3} \hat{\rho}$ where
$\tilde{\rho}$ is the matter frame conserved energy density and $\hat{\rho}$ is the Einstein frame conserved energy density such that $\hat{\rho} \neq
\hat{\rho}(\phi)$.  This leads us to the relation \be V_{eff} = V(\phi) +
A(\phi) \hat{\rho}\ee This then makes for interesting behaviour in the regime
where $\rho \rightarrow 0$ since with a properly chosen $V$ can lead to quintescence
behaviour (such as the Pebles-Ratra potential~\cite{PRDE}).  Naturally if $\rho$
becomes more prominent, then the effective minima of the potential is shifted
and the mass of the field (from $m^2 = \partial^2 V/\partial \phi^2|_{\phi=\phi_{min}}$)
becomes much heavier.  In typical chameleon models, we need $A$ to take some runaway form such that the mass of the scalar field $\phi$ becomes too heavy to detect in earth based experiments but on cosmological scales $V$ becomes the dominant contribution allowing it to act as dark energy.  Additionally in this mechanism, $V =
V(\phi)$ only, otherwise other kinds of screening are present.  Originally
$A(\phi)$ mechanisms were considered, however these have been generalised
to those with derivative screening~\cite{derivcham}, $A(\phi,X)$.  

{\sl A Toy Model.} 
We begin with the standard treatment from the T$e$V$e$S action~\cite{teves}, neglecting
however the vectorial terms (a treatment that we can justify safe in the
knowledge that the vector field does not enter into the weak field limit
of the theory, so as far as quasi-static systems see, this is the effective
theory for TeVeS),
\bea S &=& \frac{1}{2}\int \,\sqrt{-g}\, d^4 x \,\left(\frac{R}{8 \pi G} - \frac{f}{\kappa G} g^{\mu\nu}\partial_\mu \phi\partial_\nu \phi \right.
\\  \nonumber&-& \left.\frac{V(f)}{2\kappa \ell^2 G}\right) + \int \mathcal{L}_m\,\left(\tilde{g}_{\mn}, f^\alpha,\dots\right)\,\sqrt{-\tilde{g}}\,d^4 x \eea where here $\tilde{g}\mn = e^{-2\phi} g\mn$, i.e. there exists a fully conformal mapping between the gravity and matter metrics (a feature generically not true with the addition of vector fields here)


From this we find the equations of motion for the metric as the usual Einstein
equation with the bimetrically coupled matter\bea G\mn &=& 8\pi G \left(\tilde{T}\mn
+ \tau\mn\right)\\ \tilde{T}\mn &=& \frac{2}{\sqrt{-\tilde{g}}}\frac{\delta(\sqrt{-\tilde{g}}\mathcal{L}_m)}{\delta\tilde{g}\mn}\\
\tau\mn &=& \frac{f}{\kappa G} \left(\partial_\mu\phi\partial_\nu\phi
- \frac{1}{2} g\mn g^{\alpha\beta}\partial_\alpha\phi\partial_\beta\phi -
\frac{V(f)}{f\kappa\ell^2}g\mn\right)\nonumber\\ \eea additionally we have the $\phi$ field equation
\be \Del_\nu\left(\frac{f}{\kappa G}\,g^{\mu\nu}\partial_\mu \phi\right) = \tilde{T} =
\tilde{\rho} e^{-2\phi} \label{modpoisson}\ee
where $\Del_\mu$ is the covariant derivative associated with the Einstein
frame metric $g\mn$ and we source the stress energy with a pressureless perfect fluid in the matter frame. Finally for the non-dynamical scalar $f$, we find an equation of motion
of the form \be -\frac{1}{2}\frac{\partial V}{\partial f} = \kappa \ell^2 g\umn\partial_\mu \phi \partial_\nu \phi \ee where $\ell$ is a length scale related to $a_0$.  From the expansion
of (\ref{modpoisson}), \be f\square \phi = \kappa G \tilde{\rho}e^{-2\phi} - g\umn\Del_\mu f\Del_\nu\phi\label{MONDpoisson}\ee where $\tilde{\rho}$ is the matter frame density, related to the Einstein frame conserved density by
$\tilde{\rho} = A^{-3}(\phi)\hat{\rho}$.  Putting this together and rearranging \bea &\square \phi& = (f^{-1}\kappa G e^{4\phi}) \hat{\rho} - f^{-1}\underbrace{g\umn
\Del_\mu f \Del_\nu\phi}_{-2 X \partial f/\partial \phi} \\ &\Rightarrow&
= A_{,\phi}\,\hat{\rho} + V_{,\phi} \eea
where it is crucial $V = V(\phi)$ only and we will show how this is achieved
later.

\subsection{Recasting A Screening Mechanism as a Preferred Acceleration Theory}

First we need to define the order parameter $z$ \be z = \frac{|\Del \phi|}{M}\ee and consider the modified Poisson equation \be \Del_\mu (f \Del^\mu \phi) =
f\Del_\mu\Del^\mu \phi + \Del_\mu f \Del^\mu \phi = \kappa G \rho e^{-2\phi}\label{mondP}\ee such that for $z \gg 1$, $f \rightarrow 1$.  At this stage, we consider dropping
the exponential term on the source, citing that we are using a quasi-static
approximation here (see Section IIIC of~\cite{teves} for more details on this step).  

Comparing this 
with the Chameleon equation of motion \be \Del_\mu\Del^\mu \phi - V_{,\phi}
= A_{,\phi}\hat{\rho} \ee giving us a clear association between \be \frac{\partial
V}{\partial \phi} = -\frac{1}{f}\Del_\mu f\Del^\mu \phi = -\frac{1}{f}\frac{\partial f}{\partial \phi}\Del_\mu \phi \Del^\mu \phi = 2 X \frac{\partial \ln f}{\partial
\phi} \ee \be V = 2 X \ln f \Longleftrightarrow f = \exp\left(\frac{V}{2X}\right)
\ee
Recall the Peebles Ratra potential~\cite{peeblesratra} \be V = C_1^2 \frac{M_{pl}^{n+4}}{\phi^n} \ee  making \be f = \exp\left(-C_1^2 \frac{M_{pl}^{n+4}}{\phi^n|\Del \phi|^2}\right)\ee
 so this represents a more general free function in $f(\phi,z)$. 

\subsection{An $n = 0$ theory} It is worth therefore associating \be z = \frac{|\Del\phi|}{C_1 M_{pl}^{2}} = \frac{\kappa}{4\pi}\frac{|\Del\phi|}{a_0} = \frac{|\Del\phi|}{M}\ee making \be f = \exp\left(-\frac{1}{z^2}\right)\ee

Given we are entering the quasi-static regime with effective equation of
motion\be \Del\cdot(f \Del\phi) =
\kappa G \tilde{\rho}\ee we will fix \be
 \kappa  \rightarrow \frac{C_2}{2G}\ee where $C_2/2$ can be thought of as the
 limiting value of $A(\phi)$ as $\phi \rightarrow 0$, this gives us a definition for $a_0$ of \be  a_0 = C_1 C_2 M_{pl}^{4}  \ee 
These together give the relation \be f = \exp\left(-\frac{1}{z^2}\right)\ee
which obviously satisfies $f \rightarrow 1, z \gg 1$.  Such relations then
mean that the linear variable choice $U = f z$ will be in the regimes of
$U \gg 1 \rightarrow z \gg 1$ and so the bubble size can be inferred from
$|\vU|^2 \simeq 1$ \bea z^2 \simeq 1 \Rightarrow \left(\frac{\kappa}{4\pi}\right)^2
\frac{A^2 r^2 N^2}{M^2} &\simeq& 1 \nonumber \\ r^2 |\vN|^2 = \left(\frac{4 \pi M}{\kappa A}\right)^2 = \left(\frac{C_1}{C_2 A}\right)^2 &=& r_0^2\eea For the Earth-Sun SP, this takes the value $r_0 \simeq C_1/C_2 \times 10^{8}$ km.\\

\subsubsection{$z \ll 1$ - Inner Bubble Regime}\label{n=0inner}

Here the issue is that $f$ vanishes at $z=0$ and so there is no expansion
we can make here, however since we are unlikely to sample the signal {\it
exactly} at the SP, we can make do with an expansion at small $z$.  The conclusion
of which is \bea f \simeq \lim_{p\rightarrow \infty} z^p &,& z \ll 1\eea and using the standard tools to compute the form of $\vF_\phi$~\cite{bekmag,aliscaling},
we find 
\bea -\Del\phi &\simeq& M \lim_{p \rightarrow \infty}\left( C^{\frac{1}{p+1}}\frac{\vect{D}}{D^{\frac{p}{p+1}}}\left(\frac{r}{r_0}\right)^{\frac{1}{p+1}}\right)\\
&\simeq & M \,\left(F_0(\psi) \,\ve_r + G_0(\psi)\,\ve_\psi\right)\label{n=0force}\eea
where we approximate the deep inner bubble solutions with \bea F_0 &\simeq&
0.024 + 0.886 \cos 2\psi - 0.012 \cos 4\psi \\ G_0 &\simeq & -1.090 \sin
2\psi + 0.022 \sin 4\psi\eea
The tidal stresses therefore are \bea S_{yy} &\rightarrow&  \frac{C_1 M_{pl}^2}{2r}S_1(\psi)
+ \frac{C_2 M_{pl}^2 A}{2}\eea where $S(\psi)$ is computed from the
separable ansatz profile functions and the charge of variable and $A$ is
the Newtonian tidal stress at the SP.  We plot the predicted spatial variations
of the tidal stresses in Figure \ref{fig:stress} (in arbitrary units), noting both the very different radial dependence as well as the different overall
profile function.  

\begin{figure}[h!]\begin{center}
\resizebox{1\columnwidth}{!}{\includegraphics{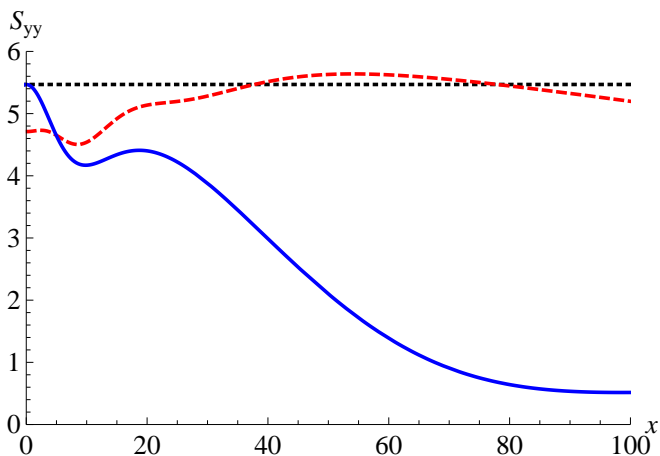}}
\caption{\label{fig:stress}{Comparing the expected spatial variation of tidal stress signals from a typical MONDian signal (red, dashed line), expected
Newtonian signal (black, dotted line) and an expected signal from a chameleon
(blue, solid line) as potential measured.  The axes here label the magnitude
of one component of the transverse tidal stress ($S_{yy}$) in arbitrary units
as well as the spatial variation along a trajectory (along the $x$ axes)
with a finite miss distance from the SP.  }}\end{center}
\end{figure}

\subsubsection{ $z \gg 1$ - Outer Bubble Regime}
In the outer bubble regime, we simply expand $f$ \be f \simeq 1 - \frac{1}{z^2}
+ \dots \ee and proceed to find solutions using the standard techniques in
these theories.  The expected tidal stresses therefore
are of the form \bea S_{yy} = \frac{C_1 M^2_{pl}}{2}S_2(\psi)r^{-2} \eea
where $S_2(\psi)$ is presented in full in~\cite{aliscaling}.  

\subsection{An $n \neq 0$ Theory}

The identification there must be made here is \bea f &=& \exp\left(-C_1^2\frac{M_{pl}^{n
+ 4}}{|\Del\phi|^2\phi^n}\right)
\\ \phi^n|\Del\phi|^2 & = &\left(\frac{2}{n+2}\right)^2|\Del\tilde{\phi}|^2\\
\tilde{\phi} &=& \phi^{1 + n/2} \\ z &=& \frac{|\Del\tilde{\phi}|}{M} \\ M &=& C_1 M_{pl}^{n/2 + 2}\left(1 + \frac{n}{2}\right)\eea
Additionally we see that the bubble boundary is modified \bea |\vU|^2 \simeq
1 &\Rightarrow& |z|^2 \simeq 1 \\ \phi^n |\Del\phi|^2 &\simeq& M^2 \\ r^{2(n+1)}
(N_r)^n |\vN|^2 &\simeq& r_0^2\eea
And as such, \bea a_0 &=& C_1 \,C_2 \,M_{pl}^{n/2 + 4}\, \left(1 + \frac{n}{2}\right)
\\ r_0 &=& \frac{2^n \,C_1}{(A \,C_2)^{n+1}}\, M_{pl}^{-3n/2} \,\left(1 + \frac{n}{2}\right)\eea
The key feature of this result is that it implies regions close to the saddle
are generically inside the modified regime.

 Here we can follow the same procedure as
in Section \ref{n=0inner}, working with an expression for $f$
 \bea f &\simeq& z^q \\ U &\simeq& z^{q+1} \nonumber \\f &\simeq& U^{\frac{q}{q+1}}\eea
 where ultimately we are taking $q \rightarrow \infty$.  Putting this together
 gives \be -\Del\tilde{\phi} \simeq M\frac{\vD}{D^{\frac{q}{q+1}}}\left(\frac{r}{r_0}\right)^{\frac{1}{q+1}}\ee which in the large $q$ limit reduces to expression for $\phi$ in (\ref{n=0force}).  However to recover the
actual force from the physical potential $\phi$, we first reduce to the potential \bea \tilde{\phi}
&=& -M F_0\, r \\ \phi &=& \tilde{\phi}^{\hat{n}} = -\left(M\,F_0\, r\right)^{\hat{n}} \\-\Del\phi &=& \tilde{M}\, r^{\hat{n}-1}\left(F_n(\psi) \ve_r + G_n(\psi)\ve_\psi\right) \\ (F_n,G_n) &=& \left((F_0)^{\hat{n}},(G_0)^{\hat{n}}\right)\\ \tilde{M} &=& C_1^{\,\hat{n}}M_{pl}^{n+4/n+2} \hat{n}^{-\hat{n}} \\ \hat{n} &=& \frac{2}{n+2}\eea which clearly reduces to (\ref{n=0force}) for $n=0$ but here generalises our result.  The corresponding
tidal stresses therefore are \bea S_{yy} &=& a_1 \,S_3(\psi) \,r^{-c} \\c &=& \frac{2n+2}{n+2} \\ a_1 &=& \frac{\tilde{M}}{2}\eea where $S_3(\psi)$ is calculated
from the change of variable and components of $\Del\phi$ (more details of
which can be found in~\cite{aliscaling}).  These results show that the tidal stresses diverge with radial exponent $1 < c < 2$.  

\begin{figure}[h!]\begin{center}
\resizebox{1\columnwidth}{!}{\includegraphics{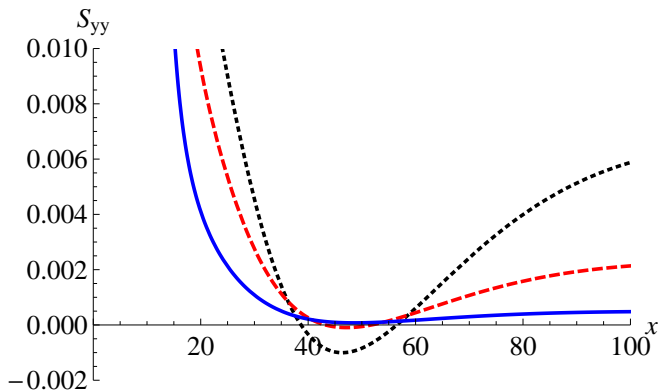}}
\caption{\label{fig:stress}{Comparing the expected spatial variation in tidal stress signals from a typical chameleon signal for $n=1$ (black, dotted line), $n=2$ (red, dashed line) and $n=5$ (blue, solid line). }}\end{center}
\end{figure}

 \subsubsection{Regime of validity}
 The next question to ask is how valid our our solutions, this points to
 the validity of the quasi-static regime used to derive our relation (\ref{mondP}).
  We see that if \be \exp[-2\phi] \simeq 1\ee then our relation \be \Del\cdot\left(f
  \Del\phi\right) = \frac{\kappa}{4\pi}\Del^2\Phi_N\ee holds, the departures
  from this signal the onset of the full chameleon mechanism and the regime
  where our results are not applicable.   As a rough estimate, we could take
  this to mean for departures of $|\delta| \sim \mathcal{O}(10^{-1})$
  \bea \delta &\simeq& 2 M F_0 r \\  C_1 \,r &\lesssim& 1.7\times10^{-10} \label{QSregime}\eea

 \subsubsection{Constraints from data}
 
 Different types of constraint are expected to hold for these models: 
 \begin{itemize}
 \item{\bf $G_N$ Renormalisation}
 It is rather bad form to let our effective gravitational constant $G$ vary
 from $G_N$ by too much, partly because it will mess up the cosmology of
 such theories.  The contribution from $\phi$ can be see in the large $z$
 limit as \be G_{eff} = G_N \left(1 + \frac{\kappa}{4\pi}\right) = G_N (1 + C_2 M_{pl}^2) \ee Thus for $|\Delta G| \lesssim 10^{-1}$~\cite{Nconstraint}, \be C_2 \lesssim 1.7 \times 10^{-10} \ee 

\item{\bf Sensitivity}
Given that we have not detected anything like a signal from a fifth force field in the Solar System, it is prudent to imagine that only with new experiments
could the possibility of detection become viable.  As a naive first consideration,
bounds on fifth forces from variations in Kepler's constant and precessions
of Mercury and other inner Solar System objects~\cite{SSconst}.  This gives
an upper bound on the size of such forces (along with the expectation that
they are ``long range'').  Thus we argue that they could be hidden within
the sensitivity of current measurements, putting a bound on $C_1$, of
the order of  \be C_1 \lesssim 10^{-15}\ee  Given this, the magnitude of
the expected tidal stresses are within the range accessible from LPF.  If $C_1$ is drastically smaller than this, this signal will unlikely to seen
above the background and noise (although for the more complicated $n \neq
0$ models with stronger divergences, this remark is subject to change).  Taking this idea from the reverse point
of view, if no signal is seen, this represents the best constraint on $C_1$
that we can make.  

If we put these together with the quasi-static requirement, this results
in \be r\simeq 1.7\times10^4\unit{m}\ee which is at similar level as the current best estimates for a SP miss with LPF. 
 \end{itemize}
 
 \subsection{Conclusions}
 
 In this work we develop a test for Chameleon screening mechanisms in the
 Solar System using the forthcoming LISA Pathfinder mission.  We recast such
 theories in the language of modified gravity theories with a preferred acceleration scale.  In doing so we present the expected tidal stresses for such theories around the gravitational SP in the Solar System, specialising to the Earth-Sun
 system.  Using a combination of analytical results and numerical suggestions
 we propose that such a test could make it possible to test such theories
 cleanly, depending upon the precise details of the models used.  In a forthcoming
 paper~\cite{alijohannes}, we will expand on our methods as well as focus on other screening
 mechanism and the prospects for observation.


\begin{acknowledgments}
The author would like to thank Johannes Noller for exceptionally useful discussions
and acknowledges the STFC, Department of Physics and Centre for Co-Curricular Studies, Imperial College for financial and other support during the various stages of this work.
\end{acknowledgments}


\bibliographystyle{ieeetr}
\bibliography{references}

\end{document}